\documentclass{ifacconf}

\counterwithin*{section}{part}

\usepackage[comma]{natbib}        
\usepackage[abs]{overpic}

\usepackage{amsmath,amssymb,graphicx,subcaption,enumerate,xcolor}

\newcommand{\cD}{\mathcal{D}}
\newcommand{\cC}{\mathcal{C}}
\newcommand{\R}{\mathbb{R}}
\newcommand{\N}{\mathbb{N}}
\newcommand{\ddt}{\tfrac{\text{d}}{\text{d} t}}

\newcommand{\dom}{\operatorname{dom}}

\DeclareMathOperator{\Span}{span}

\usepackage{cleveref}
\newtheorem{theorem}{Theorem}[section]
\newtheorem{lemma}{Lemma}[section]

\newtheorem{remark}{Remark}[section]
\newtheorem{definition}{Definition}
\newtheorem{assumption}{Assumption}
\newtheorem{problem}{Problem}
\newtheorem{proposition}{Proposition}[section]

\Crefname{theorem}{Theorem}{Theorems}
\Crefname{proposition}{Proposition}{Propositions}
\Crefname{lemma}{Lemma}{Lemmas}
\Crefname{corollary}{Corollary}{Corollaries}
\Crefname{remark}{Remark}{Remarks}
\Crefname{definition}{Definition}{Definitions}
\Crefname{assumption}{Assumption}{Assumptions} 
\Crefname{figure}{Fig.}{Figs.}

\newenvironment{smallpmatrix}%
{\left(\begin{smallmatrix}}%
{\end{smallmatrix}\right)}%

\newenvironment{smallbmatrix}%
{\left[\begin{smallmatrix}}%
{\end{smallmatrix}\right]}%

\begin{document}
\begin{frontmatter}
\title{Lyapunov based dynamic controller designs for reach-and-avoid problems\thanksref{footnoteinfo}}
\thanks[footnoteinfo]{
L.~Lanza is grateful for funding by the German Research Foundation (DFG; Project No.~471539468). 
P. Braun is partially supported by the ANU CIICADA lab and the Grant No. FA2386-24-1-4014.
}

\author[Second]{Lukas Lanza} 
\author[First]{Philipp Braun}

\address[Second]{Technische Universit\"at Ilmenau, Institute of Mathematics, Optimization-based Control Group, Ilmenau, Germany \\ (e-mail: lukas.lanza@tu-ilmenau.de).}
\address[First]{Australian National University, School of Engineering, Canberra, Australia (e-mail: philipp.braun@anu.edu.au).}

\begin{abstract}                %
Safe obstacle avoidance and target set stabilization for nonlinear systems 
using reactive feedback control is under consideration.
Based only on local information and by considering virtual dynamics, a safe 
path is 
generated online.
The control law for the virtual dynamics is combined with a feedback controller for the dynamics of interest, where Lyapunov arguments and forward invariance are used to ensure that the state of the system remains in a vicinity of the path.
To allow for discrete decisions in the avoidance controller design, the
closed-loop dynamics are formulated using the hybrid systems framework.
The results are illustrated by a numerical example for unicycle dynamics.
\end{abstract}

\end{frontmatter}

\section{Introduction}

Controller designs for 
nonlinear systems aiming for global convergence to a target set and simultaneous avoidance of unsafe sets (i.e. obstacles) have a long history of contributions from researchers and practitioner working on autonomous vehicles, including mobile robots, drones, ships, and
submersibles. 
The inherent conflict between the objectives of these reach-and-avoid problems
pose persistent challenges in controller designs, which, despite various approaches in the literature, prevent general solution concepts, solving the underlying problem for arbitrary dynamics, arbitrary unsafe sets and potentially for moving obstacles. 
This is in particular the case, if one is after feedback controllers 
with global reach-and-avoid and robustness properties, where topological obstructions demand for
discontinuous (or dynamic time-varying) controller designs \cite[Ch. 4.1]{switching_liberzon2003}, \cite{BRAUN2020109225}.

Classical approaches addressing the combined control problem with convergence and avoidance properties  dating back to the late 1980s are based on artificial potential fields, cf.~\cite{163777,8114216}. These approaches are usually restricted to fully actuated systems and use the gradient of a smooth potential field for controller designs. Since these feedback laws are Lipschitz continuous, works on artificial potential fields in general remove sets of measure zero from their performance analysis.
In recent years combinations of control barrier functions and control Lyapunov functions (see for example \cite{7782377,8796030}) have been used to address reach-and-avoid problems.  
Similar to artificial potential fields, level sets of barrier functions guarantee avoidance and control Lyapunov functions guarantee convergence.    
In this context, high-order control barrier functions (see \cite{9456981,9516971}, for example) overcome the restriction to fully actuated systems, but do not circumvent the need for discontinuous feedback laws. Results, pointing out the effect of unwanted equilibria in the context of control Lyapunov and control barrier function based controller designs can be found in \cite{BRAUN2020109225,reis2020control}, for example.

Model predictive control (MPC), see e.g.~\cite{Grune2017}, is a standard approach for constrained  nonlinear control problems and thus, naturally finds its applications in reach-and-avoid applications. 
Under the assumption of bounded unsafe sets, the optimization problem that needs to be solved at every 
time step is in general non-convex, 
complicating the performance analysis and the real-time capabilities of the MPC schemes.
Alternative approaches (including publications on MPC)
split the reach-and-avoid problem in two separate tasks: (1) path planning and (2) reference tracking or path following, cf.~\cite{Nascimento_Dórea_Gonçalves_2018,Hoy_Matveev_Savkin_2015}.

Reference governors, cf.~\cite{GARONE2017306},
constitute an additional framework with applications covering reach-and-avoid problems taking constraints directly into account.
Reference governors use an additional control 
layer generating a reference signal for the  
dynamics of interest. If designed and updated correctly, the reference signal ensures that state and input constraints are satisfied and performance goals are met. A related approach is used in 
\cite{icsleyen2022low,icsleyen2023adaptive}, for example, to design safe reach-and-avoid controllers for unicycle dynamics with 
performance metrics in terms of faster robot motion.

In this paper we introduce a related hierarchical controller design relying on a controller for fully actuated virtual dynamics whose state acts as a reference signal, and a controller for the 
system of interest, ensuring that the output 
of the system stays in a vicinity of the reference signal. Thus 
the reference signal guides the output 
to 
a target set while forward invariance of sublevel sets of 
Lyapunov functions guarantee safety and obstacle avoidance. To circumvent issues with saddle points and local minima behind obstacles, 
a discontinuous feedback law relying on the hybrid systems framework \cite{TeelBook12}
is chosen for the controller design for the virtual dynamics. 
The paper extends previous work in \cite{braun_2024}, and 
the controller design for the virtual dynamics is inspired by the construction in~\cite{berkane2019hybrid}.

The paper is structured as follows. \Cref{eq:problem_formualtion} introduces the dynamics of interest and concludes with a
formulation of the reach-and-avoid problem solved in this paper. 
\Cref{Sec:ControllerForVirtualSystem} discusses a controller design with global avoidance and global convergence properties for fully actuated linear dynamics
using the hybrid systems framework. 
By using the state of the 
linear dynamics 
as a reference, \Cref{Sec:ControllerDesign} combines the dynamics of interest with the closed-loop dynamics in \Cref{Sec:ControllerForVirtualSystem}. In particular,
the state of the linear dynamics and a Lyapunov based controller 
for nonlinear systems are used to derive a solution of the reach-and-avoid problem. 
Numerical illustrations on the example of extended unicycle dynamics are given in  \Cref{Sec:Numerics} before the paper concludes with an outlook to future work. 

\textbf{Notation.}
In this paper let ${\R_{\ge0} : = \{ a \in \R \, | \, a \ge 0\}}$ and ${\R_{>0} : = \{ a \in \R \, | \, a > 0\}}$.
For $x,y\in \R^n$ we set $\|x\| := \sqrt{x^\top x}$ and $\langle x,y \rangle := x^\top y$.
For $r \in \R_{>0}$ and $q\in \R^n$ we define $B_r(q) := \{ x \in \R^n \, | \, \|x-q\| < r \}$.
For a set $B\subset \R^n$, the closure and the boundary are denoted by $\overline{B}$, $\partial B$,
and for $x\in \R^n$ we define $\|x\|_B := \inf_{y\in B}\|x-y\|$.
For $q \in \R^n$ $\Span(q):=\{\alpha q\in \R^n| \alpha \in \R\}$ denotes its linear span, and
\begin{align}
\!\! Q^{\bullet}(q) \!: =\! \{ a \in \R^n \, | \, \langle a,a\!-\!q \rangle \!\bullet\! 0 \} \text{ for }\bullet\!\in\! \{<,\!\le,\!=,\!\ge,>\} \label{eq:Q_notation}
\end{align}
defines various sets, and in two dimensions~$Q^=(q)$ is a circle.
For representations of hybrid systems and their solutions we follow the notation~\cite{TeelBook12,sanfelice2021hybrid}. 
A solution $x:\R_{\geq 0}  \times  \N \to \R^n$ to a hybrid system evolves in the hybrid time domain and its domain is denoted by $\dom(x)\subset \R_{\geq 0} \times \N$. We write $(t,j) \rightarrow \infty$ if $t+j\rightarrow \infty$ for $(t,j)\in \dom(x)$ and $(t_1,j_1)\leq (t_2,j_2)$ if $t_1+j_1 \leq t_2+j_2$ for $(t_1,j_1),(t_2,j_2)\in \dom(x)$. Readers unfamiliar with 
$\mathcal{K}_\infty$- and $\mathcal{KL}$-functions are referred to \cite[Ch. 1.1.3]{kellett2023introduction}.

\section{Problem formulation} \label{eq:problem_formualtion}
For the controller design, we consider systems
\begin{equation} \label{eq:System}
\begin{aligned}
    \dot x(t) &= f(x(t),u(t)), \quad x(0) = x_0 \in \R^n, \\
    z(t) &= h(x(t)),
\end{aligned}
\end{equation}
where~$t\in \R_{\geq 0}$ denotes time, $x(t) \in \R^n$ is the state, $u(t) \in \R^m$ 
is the control input, and $z(t)\in \R^p$ is a subset of states to be stabilized and denoted as output of interest in the following.
Accordingly, the control objective will be formulated with respect to~$z$ but we point out that~$z$ does not necessarily represent a measured output.
The functions $f: \R^n \times \R^m \to \R^n$ and $h : \R^n \to \R^p$ are Lipschitz continuous by assumption.
We consider the set
\begin{align}
    \Gamma = \left\{ (x_e,u_e)\in \R^{n + m} \ | \ 0=f(x_e,u_e), \ u_e\in \R^m \right\} \label{eq:Gamma} 
\end{align}
defining pairs of induced equilibria.
Throughout the paper, we make use of the following assumption, which ensures that 
for any desired output of interest $z_e^*\in \R^p$, there exists $(x_e^*,u_e^*) \in \Gamma$ such that $z_e^*=h(x_e^*)$. 
This assumption allows 
us to design a controller with the property $z(t) \rightarrow z_e^* \in \R^p$ for $t\rightarrow\infty$ for the corresponding closed-loop system~\eqref{eq:System}.
\begin{assumption}[{\cite[As.~1]{braun_2024}}] \label{as:equilibrium_characterization}
For all $z_e \in \R^p$ there exist $(x_e,u_e) \in \Gamma$ such that $z_e=h(x_e)$ and a Lipschitz continuous set-valued map  $G:\R^p \rightrightarrows \Gamma$,
\begin{align*}
G(z_e) &= \{ (x_e,u_e) \in \R^{n+m} \ | \ z_e=h(x_e), \ (x_e,u_e) \in \Gamma \},
\end{align*}
mapping $z_e$ to induced equilibrium pairs $(x_e,u_e)\in\Gamma$.
 \hfill $\circ$
\end{assumption}
\Cref{as:equilibrium_characterization} ensures that $G(z_e) \neq \emptyset$ for all $z_e\in \R^p$.
For a definition of Lipschitz continuity for set-valued maps we refer to~\cite[Def. 1.4.5]{aubin2009set}.
We use $G_{x}(\cdot)$ and $G_{u}(\cdot)$ in the following to refer to the first $n$ and the last $m$ components of~$G$, i.e., 
${G(z_e)=(G_x(z_e),G_u(z_e))\subset \R^n\times \R^m}$.
In addition to~\eqref{eq:System}, we consider \textit{virtual dynamics}
\begin{align}
    \dot{\xi}(t) =  \mu(\xi(t)), \quad {\xi(0) = \xi_0 \in \R^p}, \label{eq:virtual_system}
\end{align}
with virtual state~$\xi(t)\in \R^p$ and virtual input~$\mu(\xi)\in \R^p$. 
In contrast to~\eqref{eq:System}, the virtual dynamics~\eqref{eq:virtual_system} are fully actuated.
Due to the simplicity of the latter, 
control laws guaranteeing obstacle avoidance and finite-time convergence to an output of interest $\xi_e^*\in \R^p$ can be explicitly constructed.
The state~$\xi(t)$ will serve as a reference signal for the output of interest~$z(t)$ of system~\eqref{eq:System}.
Through a Lyapunov based controller design for~\eqref{eq:System}, we ensure that~$z(t)$ remains in a vicinity of~$\xi(t)$, 
i.e., we guarantee that $\|\xi(t)-z(t)\|$ is sufficiently small for all $t \in \R_{\geq 0}$ and where sufficiently small is made precise through sublevel sets of Lyapunov functions.
Accordingly, if constructed appropriately, $\{\xi(t)\in \R^p \, | \, t \in \R_{\geq 0}\}$ defines a safe path in~$\R^p$ with corresponding safety neighborhood.
The construction of the feedback controller for~\eqref{eq:System} follows the ideas in~\cite{braun_2024}. 
In the present paper, the path~$\xi(t)$ 
is constructed online while in~\cite{braun_2024} the controller is designed to keep~$z(t)$ in the vicinity of a predefined path.

\begin{remark} \label{rem:stab_origin}
Under \Cref{as:equilibrium_characterization} we can shift the target output of interest to the origin. Thus we will focus on $z_e=0$ and $\xi_e=0$ without loss of generality. 
\hfill $\diamond$
\end{remark}
To describe unsafe sets, we consider spherical obstacles in~$\R^p$, whose properties are characterized through the following assumption.
\begin{assumption}[Obstacles and safety radii]
\label{as:obstacles}
The output-space $\R^p$ is covered with $N\in \N$ static spherical obstacles. 
For $i \in 
\{1,\ldots,N\}$ an obstacle is described through its center~$q_i \in \R^p$ 
and its radius $r_i\in \R_{>0}$ defining an unsafe set $\overline{B}_{r_i}(q_i)$.
For $\delta_i>0$ we set ${\Delta_i :=r_i +\delta_i}$ as safety radius satisfying
  $  \min_{i,j \in { \{1,\ldots,N\} },  i\neq j} \|q_i - q_j\| > \Delta_i + \Delta_j $ and
  $ { B_c(0) \cap (\cup_{i=1}^N \overline{B}_{\Delta_i}(q_i)) = \emptyset}$
for~$c\in \R_{>0}$.
\hfill $\circ$
\end{assumption}
The condition that $0 \in \R^p $ is not in the vicinity of the obstacles is necessary to ensure convergence $z(t) \rightarrow 0$ for $t\rightarrow \infty$. Under \Cref{as:obstacles} we will derive a controller addressing the following problem.

\begin{problem} \label{prob:problem_formulation}
Consider the system~\eqref{eq:System}, satisfying \Cref{as:equilibrium_characterization}, and virtual dynamics~\eqref{eq:virtual_system} together with $N\in \N$ spherical obstacles satisfying \Cref{as:obstacles}. Under these assumptions design a controller such that
\begin{enumerate}[(i)]
\item finite-time convergence of the virtual state, i.e., $\xi(t) \!\rightarrow\! 0$ for $t\rightarrow T(\xi_0)$, $T:\R^p \to \R_{\geq 0}$; 
\label{aim:path_conv_to_endpoint} 
\item obstacle avoidance of the virtual state with safety distance, i.e., \label{aim:path_avoids_obstacles}
$\xi(t) \notin \cup_{i =1}^N 
\overline{B}_{\Delta_i}(q_i)$ for all~$t \in \R_{\ge 0}$; 
\item asymptotic convergence of the output of interest, i.e., \label{aim:system_casympt_onv_to_entpoint}
$z(t) \to 0$ for $t \to \infty$; and
\item obstacle avoidance of the output of interest, i.e., \label{aim_system_avoids_obstacles}
\begin{align*}
z(t) \notin 
\cup_{i = 1}^N  \overline{B}_{r_i}(q_i) \quad \forall t \in \R{\geq 0}.
\end{align*}
\end{enumerate}
\end{problem}
A solution to Problem \ref{prob:problem_formulation}
will be derived in two steps. Namely, the design of a virtual controller in \Cref{Sec:ControllerForVirtualSystem} satisfying items~\eqref{aim:path_conv_to_endpoint} and~\eqref{aim:path_avoids_obstacles} and a controller for the original system discussed in \Cref{Sec:ControllerDesign} guaranteeing items~\eqref{aim:system_casympt_onv_to_entpoint} and~\eqref{aim_system_avoids_obstacles}. In these sections also the set of initial conditions for which items~\eqref{aim:path_conv_to_endpoint} to~\eqref{aim_system_avoids_obstacles} are satisfied are made precise.

\section{Control laws for virtual dynamics} \label{Sec:ControllerForVirtualSystem}

We focus on items \eqref{aim:path_conv_to_endpoint} and~\eqref{aim:path_avoids_obstacles} of \Cref{prob:problem_formulation} 
and derive a controller $\mu$ for the virtual dynamics \eqref{eq:virtual_system} with combined convergence and avoidance properties. First, we introduce a control law, which guarantees global finite-time stability of the origin $\xi^\star_e=0$ and, adopting ideas from~\cite{berkane2019hybrid}, a control law guaranteeing obstacle avoidance. 
In the second part, we combine the controllers to achieve global safety and almost global finite-time stability. 
As a last step, we guarantee global avoidance and stability through a switching mechanism using the hybrid systems framework in \cite{TeelBook12}.

\subsection{Stabilizing and safety controller designs} \label{Sec:ControllerDesign_one_obstacle}

Due to the simplicity of the virtual dynamics~\eqref{eq:virtual_system}, a controller design stabilizing the origin is straightforward. 
Here, we select a controller $\nu_s:\R^p \rightarrow \R^p$ of the form 
\begin{align}
    \nu_s(\xi) = - \tfrac{1}{s_c(\|\xi\|)} \xi
   \label{eq:stabilizing_controller_virtual} 
\end{align}
where the nonlinear scaling $s_c:\R_{\geq 0} \rightarrow \R_{\geq 0}$ is defined as
\begin{align}
    s_c(r) = \left\{ \begin{array}{cl}
        c^{\frac{2}{3}}  r^{\frac{1}{3}} & \text{if } r \in [0,c], \\
        r & \text{if } r \in [c,\infty),  
    \end{array} \right. \label{eq:velocity_scaling}
\end{align}
for $c\in \R_{> 0}$ constant (and corresponding to $c$ in \Cref{as:obstacles}).
By construction, the feedback law~\eqref{eq:stabilizing_controller_virtual} is Lipschitz continuous on $\R^p\backslash\{0\}$ and continuous on~$\R^p$. In particular $\nu_s(0)=0$ is well defined despite the fact that $s_c(r)^{-1} \rightarrow \infty$ for $r\rightarrow 0$.
\begin{lemma} \label{lem:FiniteTimeConvergence}
    Consider the virtual dynamics \eqref{eq:virtual_system} together with the feedback law \eqref{eq:stabilizing_controller_virtual}, \eqref{eq:velocity_scaling} for $c>0$. Then the origin of the closed-loop system is globally finite-time stable.
    \hfill $\lrcorner$
\end{lemma}

\begin{pf}
We use \cite[Thm.~10.4]{kellett2023introduction} to prove the statement.
Consider the function  $V(\xi) =\|\xi\|^\frac{2}{3}$,
which is lower and upper bounded by the $\mathcal{K}_\infty$-function $\alpha(r)=r^{\frac{2}{3}}$. 
Then for all $\xi \in \R^p$ with $\| \xi \|\leq c$, it holds that 
\begin{align*}
    \langle \nabla V(\xi), \nu_s(\xi) \rangle &= \langle \tfrac{2}{3} \|\xi\|^{-\frac{4}{3}} \xi,\nu_s(\xi)  \rangle \\
    &=\langle \tfrac{2}{3} \|\xi\|^{-\frac{4}{3}} \xi, - c^{-\frac{2}{3}} \|\xi\|^{-\frac{1}{3}} \xi \rangle \\
    &= -\tfrac{2}{3} c^{-\frac{2}{3}} \|\xi\|^{\frac{1}{3}} = -\tfrac{2}{3} c^{-\frac{2}{3}} \sqrt{V(\xi)},
 \end{align*}
 i.e., local finite-time stability of the origin can be concluded. 
 For $\xi \in \R^p$ with $\| \xi \|> c$ it holds that 
\begin{align*}
    \langle \nabla V(\xi), \nu_s(\xi) \rangle &= \langle \tfrac{2}{3} \|\xi\|^{-\frac{4}{3}} \xi,\nu_s(\xi)  \rangle 
= -\tfrac{2}{3} \| \xi\|^{\frac{1}{3}} \leq -\tfrac{2}{3} c^{\frac{1}{3}}.
 \end{align*}
Hence, the set $\overline{B}_c(0)$ is reached in finite-time and thus global finite-time stability follows. 
\hfill $\qed$
\end{pf}

\begin{remark}
For the controller~\eqref{eq:stabilizing_controller_virtual} only the direction~$-\xi$~is important to guarantee asymptotic stability of the origin. The scaling $s_c(\|\xi\|)^{-1}$  is used to have a constant velocity $\|\dot{\xi}(t)\|= \|\nu_s(\xi(t))\| = 1$ for all $\|\xi(t)\|\geq c$, and in a neighborhood around~$0$ it guarantees
finite-time convergence with a continuous control law.
\hfill $\diamond$
\end{remark}
The control law $\nu_a(\xi)=0$ for all $\xi \in \R^p$ trivially ensures
obstacle avoidance.
To achieve both, reach-and-avoid, we
design $\nu_a$ differently and
define $\nu^i_a : \R^p \setminus \{q_i\} \to \R^p$,
\begin{equation} \label{eq:avoidance_controller_i_virtual}
\begin{aligned} 
    \nu_a^i(\xi) &:= \pi(\xi-q_i) \nu_s(\xi), \quad \pi(z) :=  I - \tfrac{z z^\top}{\|z\|^2},
\end{aligned}
\end{equation}
where $I \in \R^{p \times p}$ is the identity, and $\pi : \R^p \setminus \{0\} \to \R^{p \times p}$ is an orthogonal projection, cf.~\cite{berkane2019hybrid}.
Applying~\eqref{eq:avoidance_controller_i_virtual} keeps the distance to obstacle $i\in \{1,\ldots,N\}$ 
constant as stated next.
\begin{lemma} \label{lem:AvoidanceWorks}
Consider the virtual dynamics~\eqref{eq:virtual_system} together with the feedback law~\eqref{eq:avoidance_controller_i_virtual} and let $i\in \{1,\ldots,N\}$ identify the center of an obstacle $q_i\in \R^p$.
If~$\xi_0 \neq q_i$, then for all~$t \in \R_{\ge 0}$ the solution of the
closed-loop system satisfies
$\xi(t) \in \partial B_{\|\xi_0 - q_i\|}(q_i)$ ,
i.e., the distance~$\|\xi(t)-q_i\|$ is constant.
\hfill $\lrcorner$
\end{lemma}
\begin{pf}
    Seeking a contradiction, suppose the existence of~$t^* >0$ such that $\xi(t^*) \notin
    \partial B_{\|\xi_0 - q_i\|}(q_i)$.
    Absolute continuity 
    of the solution $\xi : [0,t^*] \to \R^p$ implies the existence of 
    $t_* := \sup\{ t \in [0,t^*) \, | \, \|\xi(t) -q_i\| = \|\xi_0 - q_i\| \}$.
    Then for almost all $t \in [t_*, t^*]$ it holds that
    \begin{align*}
         \ddt \tfrac{1}{2} \| \xi(t) - q_i\|^2 & =  \langle \xi(t) - q_i,  \pi(\xi(t) - q_i) \nu_s(\xi(t)) \rangle \\
      &= \langle \pi(\xi(t)-q_i) (\xi(t) - q_i), \nu_s(\xi(t)) \rangle = 0,
    \end{align*}
    using $\pi(a) a\! =\! 0$ for all~$a \!\in\! \R^p\backslash\{0\}$.
    This leads to the contradiction $\|\xi_0 \!-\! q_i\|^2 \!=\! \| \xi(t_*) \!-\! q_i \|^2  \!\neq\! \| \xi(t^*) \!-\! q_i \|^2$. 
    \hfill
    $\qed$
\end{pf}

\subsection{Obstacle avoidance with local convergence guarantees}
\begin{figure}[t!]
    \subfloat[$\xi$ bypassing~$\overline{B}_{\Delta}(q)$ using~\eqref{eq:AvoidanceAdvanced}.\label{fig:IllustrationTwoPaths}]{%
      \begin{overpic}[width=0.475\columnwidth]{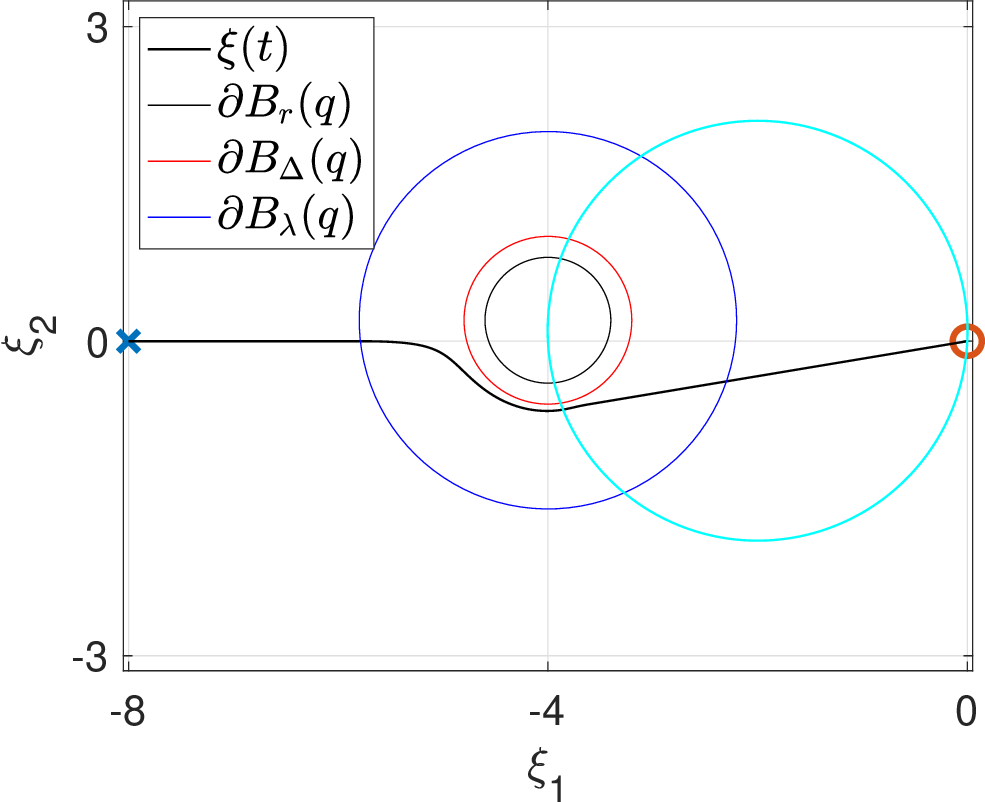}
       \put(85,41){\footnotesize{$Q^<(q)$}}
     \put(85,22){\footnotesize{$Q^>(q)$}}
      \end{overpic}
    }
    \hfill
    \subfloat[$\sigma^i(\xi(t)), \alpha^i_s(\xi(t)), \alpha^i_s(\xi(t))$. 
    \label{fig:IllustrationSigmaAlpha}]{%
      \begin{overpic}[width=0.475\columnwidth]{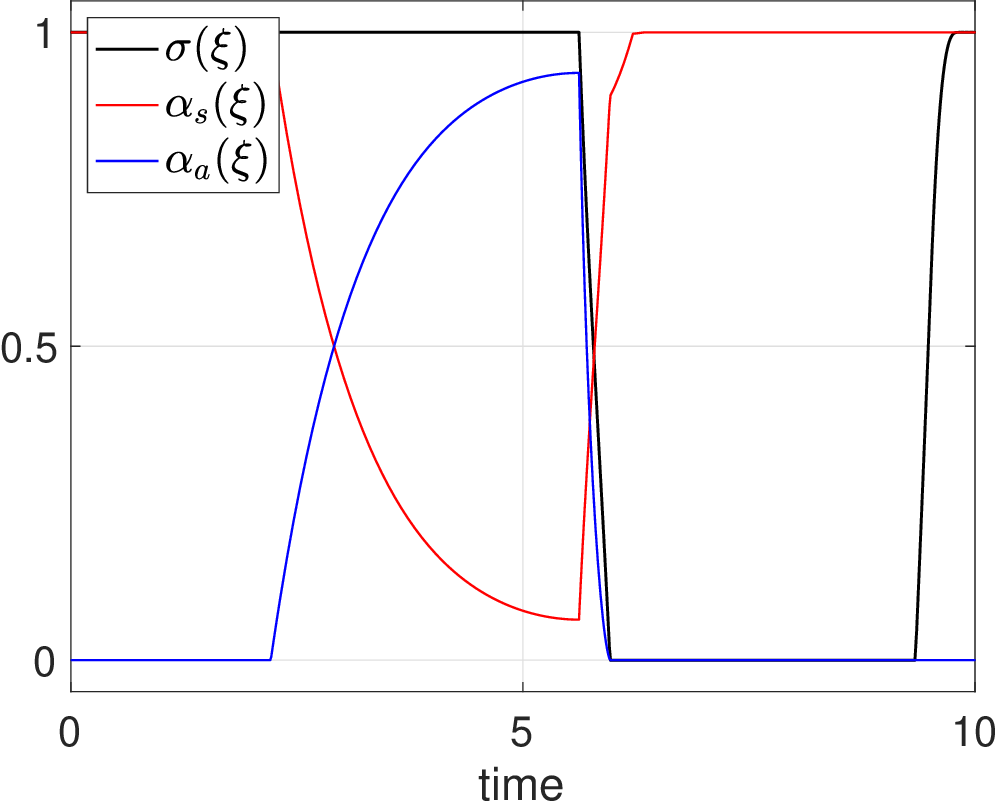}
      \end{overpic}
    }
    \vspace*{-0.5em}
    \caption{Illustration of~$\xi$ safely bypassing an obstacle~$q$ and corresponding functions $\sigma^i(\xi), \alpha^i_s(\xi)$, $\alpha^i_a(\xi)$.
    The cyan circle represents the set~$Q^=(q)$.\label{fig:dummy}}
  \end{figure}

In this section, we combine the feedback laws~\eqref{eq:stabilizing_controller_virtual} and~\eqref{eq:avoidance_controller_i_virtual}
to safely achieve the reach-and-avoid objective.
To obtain a continuous transition from the stabilizing controller $\nu_s$ to the avoidance controller $\nu_a$ when an obstacle comes in sight (cf.~\Cref{fig:IllustrationTwoPaths}), 
we introduce the following activation and switching functions based on an activation radius $\lambda_i>\Delta_i$ and based on the safety radius $\Delta_i$ introduced in \Cref{as:obstacles}. 
\begin{definition}[Activation radius] \label{def:ActivationRadii}
    Under \Cref{as:obstacles}, for $i\in \{1,\ldots,N\}$, we define an activation radius~$\lambda_i$ satisfying
    $\lambda_i \in (\Delta_i,  \min_{j \in \{1,\ldots,N\}, j \neq i} \|q_i - q_j\| - \Delta_j)$.
    \hfill $\triangleleft$
\end{definition}
Here, \Cref{as:obstacles} ensures that the intervals are non-empty and $\lambda_i$ is well-defined for all $i\in \{1,\ldots,N\}$.
With \Cref{as:obstacles} and \Cref{def:ActivationRadii}
we set the following functions.
\begin{definition}[Activation and switching functions] \label{def:activation_switch_function}
Under \Cref{as:obstacles}, let $\lambda_i$ denote an activation radius introduced in \Cref{def:ActivationRadii}.
For $i\in \{1,\ldots,N\}$ controller activation and scaling functions $\sigma^i ,\alpha_s^i,\alpha_a^i: \R^p \rightarrow [0,1]$ are defined by
    \begin{align*}
      \!\!  \sigma^i(\xi) &:= 
        \max\{ 0, \min\{ \langle \xi, \xi-q_i \rangle +1,1 \} \},  \\
      \!\!  \alpha^i_s(\xi) &:= \max\left\{0, \min \left\{ \tfrac{\|\xi - q_i\| - \Delta_i  \sigma^i(\xi) }{\lambda_i-\Delta_i}, 1 \right\}  \right\}, 
      \\
\qquad      \!\!  \alpha^i_a(\xi) &:= \sigma^i(\xi) (1-\alpha^i_s(\xi)).  
      & & \qquad \triangleleft
    \end{align*}
\end{definition}
Next we define the overall stabilizing controller with avoidance properties $\mu^i:\R^p \setminus \{ q_i \} \rightarrow \R^p$
for each obstacle by
\begin{equation} \label{eq:AvoidanceAdvanced}
\begin{aligned}
    \mu^i(\xi) = \alpha^i_s(\xi) \nu_s(\xi) + \alpha^i_a(\xi) \nu^i_a(\xi) . 
\end{aligned}
\end{equation}
Before we analyze the properties of the controller~\eqref{eq:AvoidanceAdvanced}, we provide some intuition on the functions in 
\Cref{def:activation_switch_function}.
The main idea is that during avoidance the input consists of a convex combination of the stabilizing term~$\nu_s$ and
the avoidance term~$\nu_a$.
The function~$\sigma^i$ characterizes the location of the obstacle relative to~$\xi$.
Note that for sets given in~\eqref{eq:Q_notation}, $\alpha^i_s(\xi) + \sigma^i(\xi)\alpha^i_a(\xi) = 1$ for~$\xi \in Q^{\ge}(q_i)$.
If the obstacle is between~$\xi$ and the origin, then $\xi \in Q^{\ge}(q_i)$ and~$\nu_a$ should become active when~$\xi$ enters the set $\overline{B}_{\lambda_i}(q_i)$.
If~$\xi$ reaches the set $\overline{B}_{\Delta_i}(q_i) \subset \overline{B}_{\lambda_i}(q_i)$, 
then only the avoidance term~$\nu_a$ should be active.
After the obstacle has been passed, i.e., $\xi$ is between the obstacle and the origin characterized through $\xi \in Q^{\le}(q_i)$, then only the stabilizing term~$\nu_s$ should remain active.
These phases are illustrated in \Cref{fig:IllustrationSigmaAlpha}:
while the obstacle is ``ahead'' (i.e., $\xi \in Q^{\ge}(q_i)$),  $\sigma^i(\xi) = 1$, and once the obstacle is left behind (i.e., $\xi \in Q^{\le}(q_i)$), $\sigma^i(\xi)$ decreases. 
The scaling~$\alpha_s^i(\cdot)$ ensures that~$\alpha^i_s(\xi) = 1$ while $\xi \notin B_{\lambda_i}(q_i)$, and decreases for $\xi \in B_{\lambda_i}(q_i)$.
    The scaling for the avoidance term $\alpha^i_a(\xi) = \sigma^i(\xi)(1-\alpha_s^i(\xi))$ evolves respectively.

The overall feedback law for each obstacle $\mu^i(\cdot)$ defined in~\eqref{eq:AvoidanceAdvanced} is locally Lipschitz continuous as proven below.
\begin{lemma} \label{lem:muLipschitz}
    Consider the feedback laws~\eqref{eq:stabilizing_controller_virtual} and~\eqref{eq:avoidance_controller_i_virtual} together with the functions introduced in \Cref{def:activation_switch_function}. 
    Then, under \Cref{as:obstacles}, the mapping~$\mu^i : \R^p \setminus \{q_i\} \to \R^p$ defined in~\eqref{eq:AvoidanceAdvanced}
    is continuous on~$\R^p \setminus \{q_i\}$ and locally Lipschitz continuous for all $\xi \in \R^p \setminus ( \{q_i\} \cup \{0\})$.
    \hfill $\lrcorner$
\end{lemma}
\begin{pf}
    The feedback law~$\nu_s$
    in~\eqref{eq:stabilizing_controller_virtual} is continuous on~$\R^p$ and locally Lipschitz continuous on $\R^p \backslash {\{0\}}$ by construction.
    The functions in \Cref{def:activation_switch_function} consist of multiplications and summations of bounded Lipschitz continuous functions. 
    Hence, \eqref{eq:AvoidanceAdvanced} is locally Lipschitz continuous.
    \hfill $\qed$
\end{pf}
With continuity properties of \eqref{eq:AvoidanceAdvanced} derived, we 
state reach-and-avoid
properties of the closed loop virtual dynamics~\eqref{eq:virtual_system},~\eqref{eq:AvoidanceAdvanced}. 
Here, we focus on a single obstacle before we extend the result to multiple obstacles in \Cref{Sec:HybridFormulation}.

\begin{theorem}  \label{thm:SimpleCombinedFeedback}
     Consider dynamics~\eqref{eq:virtual_system}  together with the feedback law \eqref{eq:AvoidanceAdvanced} and let \Cref{as:obstacles} be satisfied for $N=1$.
     Then, for all  $\xi_0 \in \R^p\backslash {B}_{\Delta_i}(q_i)$ it holds that
     \begin{align} \label{assertion:xi_safe_simple}
         \xi(t) \notin {B}_{\Delta_i}(q_i) \quad \forall \ t \in \R_{\geq 0}.
     \end{align}
     Moreover, for all $\xi_0 \in \R^p\setminus ( {B}_{\Delta_i}(q_i) \cup (\Span(q_i) \cap Q^{\ge}(q_i)) )$ there exists $T(\xi_0) \in [0,\infty)$ such that 
     \begin{align} \label{assertion:xi_stable_simple}
     \xi(t)=0  \qquad \forall t \ \geq T(\xi_0),
     \end{align}
     i.e., finite-time convergence to the origin.
\hfill $\lrcorner$
\end{theorem}
\begin{pf}
    If the straight line between~$\xi_0$ and the origin does not intersect with the safety ball around the obstacle, i.e., 
    if 
    $ \{b \xi_0\} \cap B_{\Delta_i}(q_i) = \emptyset $ for all~$b \in [0,1]$,
    then it holds that $\mu^i(\xi(t))= \nu_s(\xi(t))$ by construction of the functions $\alpha_s^i$ and $\alpha_a^i$ in Definition \ref{def:activation_switch_function}. Hence the statement follows from \Cref{lem:FiniteTimeConvergence}.
    We can thus focus on $ \{b \xi_0\} \cap B_{\Delta_i}(q_i) \neq \emptyset $ for some $b \in (0,1)$.
    Moreover, since only the stabilizing term~$\nu_s(\xi)$ is active if~$\xi \notin \overline{B}_{\lambda_i}(q_i)$, 
    we can further restrict our analysis to $\xi \in \overline{B}_{\lambda_i}(q_i)$. 
    We consider two phases:
    $\xi(t) \in Q^{\ge}(q_i)$ and $\xi(t) \in Q^{\le}(q_i)$.
    First, we consider ${\xi(t) \in Q^{\ge}(q_i)}$.
    Since $\alpha^i_a(\xi) = 1$ (i.e., only~$\nu_a$ is active) if, and only if, $\xi \in \partial B_{\Delta_i}(q_i)$,
    \eqref{assertion:xi_safe_simple} follows from \Cref{lem:AvoidanceWorks} for all~$t \ge 0$ such that~$\xi(t) \in Q^{\ge }(q_i)$.
    Next, consider $\xi(t) \in Q^{\le}(q_i)$.
    Suppose the existence of~$t^* > 0$ such that~$\xi(t^*) \in B_{\Delta_i}(q_i)$.
    As before, we infer the existence of
    $t_* := \max\{ t\in [0,t^*) \ | \ \xi(t) \in \partial B_{\Delta_i}(q_i) \}$.
    For almost all~$t \in [t_*,t^*]$ it holds that
    \begin{equation*}
        \begin{aligned}
          s_c(\|\xi(t)\|) \ddt \tfrac{1}{2} \| \xi(t) \!-\! q_i\|^2 \!&= \langle \xi(t) - q_i, \alpha^i_s(\xi(t)) \nu_s(\xi(t)) \rangle \\
           & =  - \alpha^i_s(\xi(t)) \langle \xi(t) - q_i, \xi(t) \rangle 
            \ge 0,
        \end{aligned}
    \end{equation*}
    and the inequality follows from the fact that
    $\xi(t) \in Q^{\le}(q_i)$ and $\alpha_s(\xi)\geq 0$ according to its definition.
    Integration yields $\Delta_i > \|\xi(t^*) - q_i\| \ge \|\xi(t_*) - q_i\| = \Delta_i$, a contradiction.
    Hence~\eqref{assertion:xi_safe_simple} is true for all~$t \ge 0$.
    To show~\eqref{assertion:xi_stable_simple}, we assume $\xi_0 \in \R^p\setminus ( {B}_{\Delta_i}(q_i) \cup (\Span(q_i) \cap Q^{\ge}(q_i)) )$ and calculate\footnote{Time arguments are omitted in the 
    calculations 
    to shorten expressions.}
    \begin{equation*}
        \begin{aligned}
       & s_c(\|\xi\|) \ddt \tfrac{1}{2} \| \xi\|^2 
         = - \langle \xi, \alpha^i_s(\xi) \xi + (1-\alpha^i_s(\xi)) \pi(\xi - q_i) \xi \rangle \\
        &= - \hspace{-0.05cm} \left( \alpha^i_s(\xi) \|\xi\|^2 \hspace{-0.05cm}+\hspace{-0.05cm} (1\hspace{-0.05cm}-\hspace{-0.05cm}\alpha^i_s(\xi)) \|\xi\|^2 \right) \hspace{-0.05cm} + \hspace{-0.05cm} (1\hspace{-0.05cm}-\hspace{-0.05cm}\alpha^i_s(\xi)) \tfrac{|\langle \xi, \xi-q_i\rangle|^2}{\|\xi-q_i\|^2} \\
        &\le - \|\xi\|^2 ( 1 - (1-\alpha^i_s(\xi)) \cos(\gamma(\xi))^2 ) \\
        & = - \|\xi\|^2 \left(\sin(\gamma(\xi))^2 + \alpha^i_s(\xi) \!\cos(\gamma(\xi))^2 \right)  
         < 0
        \end{aligned}
    \end{equation*}
    where~$\gamma(\xi)$ is the angle between~$\xi$ and~$\xi - q_i$.
    The strict inequality in the last line is implied by the fact that 
        $\sin(\gamma(\xi))^2 + \alpha^i_s(\xi) \cos(\gamma(\xi))^2 = 0$ if, and only if, 
        ${\xi \in \partial B_{\Delta_i}(q_i) \cap \Span(q_i) \cap Q^{\ge}(q_i)}$.
        The latter is excluded by assumption.
        Then, \Cref{lem:FiniteTimeConvergence} shows~\eqref{assertion:xi_stable_simple}.
        \hfill $\qed$
\end{pf}
As a next step we consider multiple obstacles and define the multi obstacle feedback law 
$\mu:\R^p \setminus \cup_{i=1}^N \{q_i\} \to \R^p$ 
\begin{align} \label{eq:virtual_system_RHS}
    \mu(\xi) = \textstyle \prod_{i =1}^N \alpha^i_s(\xi) \nu_s(\xi) + \sum_{i = 1}^N  \alpha^i_a(\xi) \nu^i_a(\xi)
\end{align}
as an extension of \eqref{eq:AvoidanceAdvanced}.
By construction~$\mu(\cdot)$ is continuous on~$\R^p$ and according to \Cref{lem:muLipschitz} locally Lipschitz continuous on $\R^p \setminus (\cup_{i=1}^N \{q_i\} \cup \{0\})$.
\begin{remark}
 The feedback law~$\mu(\xi)$ in~\eqref{eq:virtual_system_RHS} formally depends on all obstacles.
 However, since the activation and scaling functions~$\sigma^i(\xi)$, $\alpha^i_s(\xi)$ take only local information, 
 an implementation of~\eqref{eq:virtual_system_RHS} does not require knowledge of the total number of obstacles and their global positions.
 \hfill $\diamond$
\end{remark}
\begin{lemma}  \label{lem:SimpleCombinedFeedback_2}
     Consider the closed-loop dynamics~\eqref{eq:virtual_system},~\eqref{eq:AvoidanceAdvanced} under \Cref{as:obstacles}.
     Then, for all  $\xi_0 \in \R^p\setminus (\cup_{i =1}^N {B}_{\Delta_i}(q_i))$ it holds that
     $\xi(t) \notin \cup_{i =1}^N {B}_{\Delta_i}(q_i)$ for all~$t \in \R_{\ge 0}$.
     Moreover, the norm $\|\xi(\cdot)\|$ is monotonically decreasing.
\hfill $\lrcorner$
\end{lemma}
\begin{pf}
    Seeking a contradiction, suppose the existence of $t^* > 0$ such that $\xi(t^*) \in B_{\Delta_j}(q_j)$ for some index $j \in \{1,\ldots,N\}$.
    Again, by absolute continuity, there exists
    ${t_* := \sup \{ t \in [0,t^*) \, | \, \|\xi(t) -q_j\| = \Delta_j \}}$.
    Since the activation radius in \Cref{def:ActivationRadii} satisfies
    $\lambda_j < \| q_i - q_j\|$ for all $i \in \{1,\ldots,N \}\setminus\{j\}$, 
    we have $\xi(t_*) \notin \partial {B_{\lambda_i}(q_i)}$ for all $i \in \{1,\ldots,N\}\setminus\{j\}$, and thus
    \begin{align*}
        \mu(\xi(t_*)) = \mu^j(\xi(t_*)) \!=\! \nu^j_a(\xi(t_*)) 
        = \pi(\xi(t_*) - q_j) \nu_s(\xi(t_*)) .
    \end{align*}
    Following the proof of \Cref{lem:AvoidanceWorks}, we arrive at the same contradiction.
    Hence, for all~$i \in \{1,\ldots,N\}$ we conclude $\xi(t) \notin B_{\Delta_i}(q_i)$ for $t \ge 0$.
    Similar to the proof of \Cref{thm:SimpleCombinedFeedback}, we have $\ddt \|\xi(t)\|^2 \le 0$ by construction of~$\mu(\xi)$ in~\eqref{eq:virtual_system_RHS}.
    \hfill $\qed$
\end{pf}
Note that~\eqref{eq:virtual_system_RHS} yields
$\dot \xi(t) = \mu(\xi(t)) =  0$ in the case that $\xi(t)  \in \partial B_{\Delta_i}(q_i) \cap \Span(q_i) \cap Q^{\ge}(q_i)$, i.e., if obstacle $i \in \{1,\ldots,N\}$ located at~$q_i$ is on the line connecting~$\xi(t)$ and~$0 \in \R^p$ then the dynamics stop at ${\xi(t) \in \partial B_{\Delta_i}(q_i)}$.
Therefore, excluding the direct line from the initial conditions is crucial in \Cref{thm:SimpleCombinedFeedback}.
If multiple obstacles are present it is not sufficient to exclude a set of measure zero to ensure convergence to the origin. 
Depending on the locations of the obstacles, avoiding obstacle $i \in \{1,\ldots,N\}$ can lead
to ${\xi(\hat t) \in \partial B_{\Delta_j}(q_j) \cap \Span(q_j) \cap Q^{\ge}(q_j)}$ for some ${j \in \{1,\ldots,N\}}$ and some~$\hat t > 0$.
Therefore, initial conditions from an arbitrarily large set can lead to~$\dot \xi(t) = 0$, and only monotonicity (of the norm) but not convergence to the origin can be concluded through the controller design~\eqref{eq:virtual_system_RHS}.

\subsection{Augmented avoidance controller with global convergence} \label{Sec:HybridFormulation}
To achieve global reach-and-avoid properties,
we design a discontinuous feedback law with hysteresis region using the hybrid systems formalism \cite{TeelBook12}. As indicated in \Cref{thm:SimpleCombinedFeedback}, a discrete decision around $\operatorname{span}(q_i)$ is necessary to ensure that $\xi(t)$ does not stop.
We augment the state $\xi\in \R^p$ with an additional logic state $\rho\in \{0,1\}$, and focus on
\begin{align}
    \left[ 
    \begin{smallmatrix}
         \xi  \\
         \rho 
    \end{smallmatrix}
    \right] \in \mathbb{X} := \R^p\backslash \cup_{i =1}^N B_{\Delta_i}(q_i) \times \{0,1\}. \label{eq:def_X}
\end{align}
The closed set $\mathbb{X}$ defines the state space of interest excluding the unsafe sets around the obstacles.
To derive the data $(\mathcal{C},F,\mathcal{D},G)$ of the hybrid system, we focus on the flow and the jump sets and define
\begin{align*}
            {\Phi}(\theta,q,\eta) &:= \left\{ z \in \overline{B}_{\eta}(q) \, \left| \,  \tfrac{ \langle z-q,{q}\rangle }{\|z-q\| \|{q}\|} \ge  \cos(\theta) \right. \right\} ,
\end{align*}
characterized through parameters $\theta \in (0,\tfrac{\pi}{4})$, $q \in \R^p$ 
and {$\eta\in \R_{>0}$}. 
The set ${\Phi}(\theta,q,\eta)$
defines a cone-like set with aperture~$2\theta$, vertex at~$q$ and axis parallel to the vector~${q}$ 
but restricted to the sphere $\overline{B}_{\eta}(q)$. 
As a second step, for each obstacle we define the sets
\begin{equation} \label{eq:def_M}
\begin{aligned}  
    M_1^i &:=  
    {\Phi}(\theta_1^i,q_i,\lambda_i) \setminus B_{\Delta_i}(q_i) , \\
    M_0^i &:= \overline{\R^p \setminus 
    {\Phi}(\theta_0^i,q_i,\lambda_i+\varepsilon)} \setminus B_{\Delta_i}(q_i), 
\end{aligned}
\end{equation}
with $0 < \eta_1^i < \eta_0^i$ and $\varepsilon>0$.
The sets~$M_1^i,M_0^i$ are illustrated in \Cref{fig:Simulation_UniCycle} in \Cref{Sec:Numerics}.
Next, for~$a\in\{0,1\}$ we define the local and global jump sets $\cD^i_0,\cD^i_1$ and~$\cD$ by
\begin{align*}
 \cD_{a}^i \!:=\! \left\{ \left. \! \begin{smallbmatrix}
        \xi \\ \rho
    \end{smallbmatrix} \!\in\! \mathbb{X}
    \, \right\vert \, \xi \! \in \! M_{a}^i, \rho =1\!-\!a   
    \right\}, \ 
    \cD \!:=\! \cup_{i =1}^N( \cD_0^i \cup \cD_1^i) , 
\end{align*}
and the flow set $\cC := \overline{\mathbb{X} \backslash \cD}$.
As a controller, we extend~\eqref{eq:virtual_system_RHS}
\begin{subequations} \label{eq:HybridFeedbackLaw}
    \begin{align}
        \bar \mu(\xi,\rho) := & \textstyle \prod_{i = 1}^N \alpha^i_s(\xi) \nu_s(\xi) \label{eq:HybridStabilizingTerm} \\ 
        & 
        \textstyle + (1-\rho) \sum_{i =1}^N  \alpha^i_a(\xi) \nu_a(\xi) \label{eq:ControllerNomalMode} \\
        & 
        \textstyle + \rho \sum_{i =1}^N \alpha^i_a(\xi) \pi(\xi-q_i) \bar q_i, \label{eq:ControllerHybridMode}
    \end{align}
\end{subequations}
with $\bar q_i \in \ker(\xi-q_i)\setminus \{0\}$.
\begin{remark} \label{rem:ChoosingBarq}
Using~$\bar q \in \ker(\xi-q_i)$ in~\eqref{eq:ControllerHybridMode} is  \textit{one possible} choice of defining~$\bar \mu(\xi,\rho)$ in a region where the control law \eqref{eq:virtual_system_RHS} may be equal to zero. 
This ambiguity opens up the possibility to implement certain conventions, e.g., road traffic regulations, into the controller.
Since $\Span(q_i)$ is a set of measure zero, the parameters $\theta_1^i$ and $\theta_0^i$  in \eqref{eq:def_M} can simply be selected based on the numerical precision in the controller. For illustration purposes, we use large values in the numerical simulations in Section \ref{Sec:Numerics}.
\hfill $\diamond$
\end{remark}
The flow map and the jump map are defined as
\begin{equation} \label{eq:HybridPath}
    \begin{aligned}
        \left[\begin{smallmatrix}
            \dot \xi \\ \dot \rho
        \end{smallmatrix}\right]
        =F(\xi,\rho)&:= 
        \left[\begin{smallmatrix}
            \bar{\mu}(\xi,\rho) \\ 0
        \end{smallmatrix} \right], 
        \ \left[\begin{smallmatrix}
            \xi \\ \rho
        \end{smallmatrix}  \right] \in \mathcal{C}, \\
               \left[ \begin{smallmatrix}
            \xi^+ \\  \rho^+
        \end{smallmatrix} \right]
        =G(\xi,\rho)&:= 
       \left[ \begin{smallmatrix}
            \xi \\ 1-\rho
        \end{smallmatrix} \right], 
         \
        \ \left[ \begin{smallmatrix} 
            \xi \\ \rho
        \end{smallmatrix} \right]   \in \cD.
    \end{aligned}
\end{equation}
To take account of the fact that~\eqref{eq:HybridPath} is a hybrid system, we refer to it using its data~$(\cC,F,\cD,G)$.
Before we state the next result we emphasize that it is reasonable to consider the evolution of~$\xi$ in the
continuous time domain since~$\xi^+ = \xi$.
\begin{lemma} \label{lem:ExistenceSolution}
For all initial conditions $(\xi_0, \rho_0) \in \cC \cup \cD$
the closed-loop hybrid system $(\cC,F,\cD,G)$ has a nontrivial solution in the sense of~\cite[Def.~2.5]{TeelBook12}.
Moreover, all solutions are complete
and for any solution~$(\xi,\rho)$ we have $\sup_{t} \dom(\xi) = \infty$.
\hfill $\lrcorner$
\end{lemma}
\begin{pf}
The data~$(\cC,F,\cD,G)$ satisfy the conditions of \cite[Prop.~2.34]{sanfelice2021hybrid} which yields the existence of maximal solutions and that
all maximal solutions are complete, and
every maximal solution is not Zeno.
\hfill $\qed$
\end{pf}
Now we formulate our main result in terms of the virtual dynamics~\eqref{eq:virtual_system},
providing a solution to \Cref{prob:problem_formulation} items \eqref{aim:path_conv_to_endpoint} and \eqref{aim:path_avoids_obstacles} for all initial conditions $(\xi_0, \rho_0) \in \mathbb{X}$.
\begin{theorem} \label{Thm:PathGlobal}
Let \Cref{as:obstacles} be satisfied and consider the hybrid dynamics~\eqref{eq:HybridPath} defined through the data $(\cC,F,\cD,G)$.
For any initial condition $(\xi_0,\rho_0) \in \cC \cup \cD$ all solutions of the hybrid dynamics 
satisfy
\begin{equation*}
     \xi(t,j) \notin \cup_{i =1}^N B_{\Delta_i} \qquad \forall \, (t,j) \in \dom(\xi) 
\end{equation*}
    and there exists $T=T(\xi_0,\rho_0)\in[0,\infty)$ such that
    \begin{equation*} 
         \left[\begin{smallmatrix}\xi(t,j) \\ \rho(t,j) \end{smallmatrix} \right] = 0 \ \ \forall (t,j) \in \{(t,j)\in \dom(\xi) \, | \,  t+j\geq T\},
    \end{equation*}
    i.e., finite-time convergence of the hybrid state.
    \hfill $\lrcorner$
\end{theorem}%
\begin{pf}
    If all solutions satisfy $\rho(t,j) =0$ for all  $(t,j)\in \dom(\rho)$
    or $(\xi_0,\rho_0) \in \cup_{i \in I} M_0^i \times \{1\}$ and $\rho(t,j) =0$ for all $(t,j)\in \dom(\rho)$,
    then nothing is left to show thanks to \Cref{thm:SimpleCombinedFeedback} and \Cref{lem:SimpleCombinedFeedback_2}.
    We separately consider the two cases $(\xi(t_0, j_0-1),\rho(t_0, j_1-1)) \in \cD^i_0$ for some $(t_0,j_0-1) \in \dom(\xi)$,
    and $(\xi(t_1, j_1-1),\rho(t_1, j_1-1)) \in \cD^i_1$ for some $(t_1,j_1-1) \in \dom(\xi)$.
    Let $(\xi(t_0, j_0-1),\rho(t_0, j_0-1)) \in \cD^i_0$. 
    Since $G(\cD) \cap \cD = \emptyset$ by construction, we have that
    ${t_0 < t_0' := \sup \{ t \ge 0 \, | \, (\xi(t, j_0),\rho(t, j_0)) \in \cC \}}$ and
    $(\xi(t, j_0),\rho(t, j_0)) \in M_0^i \times \{0\} \subset \cC $ for all ${t \in [t_0,t_0')}$.
    \Cref{lem:SimpleCombinedFeedback_2} yields $\xi(t,j_0) \notin \cup_{i =1}^N B_{\Delta_i}(q_i)$ for all ${t \in [t_0,t_0')}$.
    Moreover, for almost all $t \in [t_0,t_0')$ we have that $\ddt \|\xi(t,j_0) - q_i\| \ge 0$, 
    which implies that every obstacle is ``visited'' at most once.
    Let $(\xi(t_1, j_1-1),\rho(t_1, j_1-1)) \in \cD^i_1$.
    Again, we infer the existence of
    $t_1 < t_1' := \sup \{ t \ge 0 \, | \, (\xi(t, j_1),\rho(t, j_1)) \in \cC \}$ such that
    $(\xi(t, j_1),\rho(t, j_1)) \in M_1^i \times \{1\} \subset \cC $ for all ${t \in [t_1,t_1')}$.
    Moreover, $\xi(t, j_1) \in Q^{\ge}(q_i)$ for all ${t \in [t_1,t_1')}$.
    The Cauchy-Schwarz inequality yields that for almost all ${t \in [t_1, t_1')}$
    we have
    $
        \langle \dot \xi(t,k), \bar q_i \rangle = \langle \pi(\xi(t,j_1) - q_i) \bar q_i, \bar q_i \rangle >  \sin(\gamma(\xi))^2 \|\bar q_i\|^2,
    $
    where~$\gamma(\xi) \neq 0$ is the angle between $\xi(t,j_1)$ and~$\bar q_i$, which is bounded away from zero since~$\xi(t,j_1) \in Q^{\ge}(q_i)$ for all $t \in [t_1,t_1')$.
    Thus, the dynamics~\eqref{eq:HybridPath} evolve in direction~$\bar q_i$ with $\dot \xi$ proportional to~$\sin(\gamma)^{2}$.
    Since~$M_1^i$ is bounded, this implies~$t_1' < \infty$, and $(\xi(t_1',j_1),\rho(t_1',j_1)) \in \cD_0^i$.
    The latter means that the input switches back from~\eqref{eq:HybridStabilizingTerm}+\eqref{eq:ControllerHybridMode} to~\eqref{eq:HybridStabilizingTerm}+\eqref{eq:ControllerNomalMode} at $t=t_1'$.
    Since the number of obstacles is finite, there is~$\bar t > 0$ and some $J = J(\xi_0,\rho_0) \in \N$ such that
    $\xi(\bar t,J) \in B_c(0)$, with~$c>0$ from~\eqref{eq:velocity_scaling}.
    Thus, $\bar \mu(\xi(t,J),\rho(t,J)) = \nu_s(\xi(t,J))$ for all~$t \ge \bar t$.
    In particular, $\xi(t,J) \in M_0^i$ for~$t \ge \bar t$.
    Then, \Cref{lem:FiniteTimeConvergence} yields $\xi(t,J) \to 0$ as~$t \to T$.
    Moreover, since $0 \notin \cup_{i=1}^N Q^{>}(q_i)$, it holds~$\rho(T,J) = 0$.
    \hfill $\qed$
    \end{pf}

\section{Lyapunov based feedback law} \label{Sec:ControllerDesign}
In this section we combine the controller derived in \Cref{Sec:ControllerForVirtualSystem}
with a controller for the system~\eqref{eq:System}. The construction relies on ideas in \cite{braun_2024,braun_2024a}. 
Here, instead of staying in the vicinity of a predefined path, we use the state of the virtual dynamics~\eqref{eq:virtual_system} as a reference for the output of interest~$z$.

\begin{assumption}[\mbox{\cite[As. 2]{braun_2024}}] \label{as:lyapunov_function}
    There exist $\alpha_1,\alpha_2,\alpha_3 \in \mathcal{K}_\infty$, $V_\cdot(\cdot):\R^n \times \R^p \rightarrow \R_{\geq 0}$, $(x,z_e) \mapsto V_{z_e}(x)$, continuously differentiable with respect to $x$ and locally Lipschitz continuous with respect to $(x,z_e)$, and a locally Lipschitz continuous feedback law $u_{\cdot}(\cdot):\R^n \times \R^p \rightarrow \R^m$ such that  
the solutions of \eqref{eq:System} are forward complete for all $(x,z_e)\in \R^n \times \R^p$
    and 
    \begin{align}
     \alpha_1(\|x\|_{G_x(z_e)}) \leq V_{z_e}(x)  &\leq \alpha_2(\|x\|_{G_x(z_e)}) \label{eq:lyapunov_condition1} \\
     \langle \nabla V_{z_e} (x),f(x,u_{z_e}(x)) \rangle &\leq -\alpha_3(V_{z_e}(x)) \label{eq:lyapunov_condition}
    \end{align}
holds    
    for all $(x,z_e)\in \R^n \times \R^p$.
    \hfill $\circ$
\end{assumption}
Under \Cref{as:lyapunov_function} the following result is obtained.
\begin{proposition}[\mbox{\cite[Prop. 1]{braun_2024}}] \label{prop:lasalle_yoshizawa}
Consider the dynamical system \eqref{eq:System} and let $z_e\in \R^p$ be fixed. If \Cref{as:equilibrium_characterization,as:lyapunov_function}
are satisfied, then there exists a feedback law $u_{z_e}:\R^n \rightarrow \R^m$ and $\beta \in \mathcal{KL}$  such that
\begin{align}
\|x(t)\|_{G_x(z_e)} \leq \beta(\|x(0)\|_{G_x(z_e)},t), \ \  \forall \ t\in \R_{\geq 0} \label{eq:convergence_z} 
\end{align}
for all $x_0\in \R^n$. Moreover, forward invariance
\begin{align*}
x(t) \in \{ x\in \R^n | V_{z_e}(x) \leq V_{z_e}(x_0)\}, \ \  \forall \ t\in \R_{\geq 0}
\end{align*}
is satisfied for all $x_0\in \R^n$.
\hfill $\lrcorner$
\end{proposition}
\Cref{prop:lasalle_yoshizawa} guarantees that under \Cref{as:equilibrium_characterization,as:lyapunov_function} any output of interest can be asymptotically stabilized while remaining in the sublevel set of a Lyapunov function depending on the initial condition. 

To obtain safe sublevel sets of the Lyapunov function, we consider a continuous function $d:\R^p \rightarrow \R$ satisfying 
\begin{align}
    d(z_e) \leq \inf_{\substack{z\in \cup_{i=1}^N \overline{B}_{r_i}(q_i)\\ z=h(x), \ x\in \R^n}} V_{z_e}(x) - \tilde \varepsilon 
    \label{eq:def_d}
\end{align}
for $\tilde{\varepsilon}>0$ arbitrary.
The selection~\eqref{eq:def_d} ensures that for all ${x\in \{x\in \R^n \ | \ V_{z_e}(x) \leq d(z_e)\}}$ it holds that
$h(x)\notin \cup_{i=1}^N \overline{B}_{r_i}(q_i)$.
With this selection, for $\ell>0$ constant, we define the overall closed-loop dynamics
\begin{align} 
\begin{split}
        \left[\begin{smallmatrix}
            \dot{x} \\
            \dot \zeta  \\ 
            \dot \rho 
        \end{smallmatrix} \right]
        & \hspace{-0.05cm} := \hspace{-0.05cm} \left[ 
        \begin{smallmatrix}
          f(x,u_{\zeta}(x)) \\
          \max\{0, \ell (d(\zeta)\!-\!V_{\zeta}(x))\}  \bar{\mu}(\zeta,\rho) \\ 
          0 
        \end{smallmatrix} \right], 
        \left[ \begin{smallmatrix}
            x \\
            \zeta \\ 
            \rho 
        \end{smallmatrix} \right]  \in  \R^n  \times   \mathcal{C}, \\
            \left[    \begin{smallmatrix}
            x^+ \\
            \zeta^+ \\  
            \rho^+ 
        \end{smallmatrix} \right]
        & \hspace{-0.05cm} := \hspace{-0.05cm} \left[ 
        \begin{smallmatrix}
           x \\
            \zeta \\
             1-\rho \\
        \end{smallmatrix} \right], 
         \
        \ \left[\begin{smallmatrix} 
            x \\
            \zeta \\ 
            \rho 
        \end{smallmatrix} \right]   \in \R^n \times \cD ,
\end{split} \label{eq:overall_hybrid_dynamics}
\end{align}
augmenting the virtual dynamics \eqref{eq:HybridPath} with the plant dynamics \eqref{eq:System} and the controller defined in \Cref{as:lyapunov_function}. 
\begin{remark}\label{rem:scaling}
Note that  $\max\{0, \ell (d(\zeta)-V_{\zeta}(x))\}$ denotes a non-negative scaling in~\eqref{eq:overall_hybrid_dynamics} that can speed up or slow down the $\xi$-dynamics in~\eqref{eq:HybridPath}.
The scaling does not change the trajectory in terms of a path in $\R^p$. To differentiate the $\xi$-dynamics in~\eqref{eq:HybridPath} and the scaled dynamics in~\eqref{eq:overall_hybrid_dynamics} we use~$\zeta$ instead of~$\xi$ in this section. 
\hfill $\diamond$
\end{remark}
\begin{theorem} \label{thm:overall_convergence_avoidance}
Let \Crefrange{as:equilibrium_characterization}{as:lyapunov_function}
be satisfied and let $\ell>0$. 
    Consider~\eqref{eq:overall_hybrid_dynamics} with $V_\cdot(\cdot)$ and $u_\cdot(\cdot)$ defined in \Cref{as:lyapunov_function},  
    $\overline{\mu}(\cdot)$ defined in~\eqref{eq:HybridFeedbackLaw} and $d(\cdot)$ defined in \eqref{eq:def_d} for $\tilde \varepsilon>0$ such that $d(z)>0$ for all $z \in \R^p \backslash \cup_{i=1}^N \overline{B}_{r_i}(\Delta_i) $.
    Then the following properties are satisfied.

\noindent \underline{(i)} For all $(x_0,\zeta_0,\rho_0) \in \R^n \times \mathbb{X}$ all solutions of \eqref{eq:overall_hybrid_dynamics} are forward complete in the continuous time argument $t\in \R_{\geq 0}$.

\noindent \underline{(ii)} Let $(x(\cdot,\cdot),\zeta(\cdot,\cdot),\rho(\cdot,\cdot))$ denote a solution to~\eqref{eq:overall_hybrid_dynamics} corresponding to an arbitrary initial condition $(x_0,\xi_0,\rho_0) \in \R^n \times \mathbb{X}$ with $d(\zeta_0)-V_{\zeta_0}(x_0)\geq 0$. 
Then $d(\zeta(t,j))-V_{\zeta(t,j)}(x(t,j))\geq 0$ for all $(t,j) \in \operatorname{dom}(x)$ and
    \begin{align} \label{eq:avoidance_property}
      \! \! \!  z(t,j)\!=\!h(x(t,j)) \!\in\! \R^p \!\setminus\! \cup_{i=1}^N \overline{B}_{r_i}(q_i) \, \forall (t,j) \!\in\! \dom(x). \! 
    \end{align}

\noindent \underline{(iii)} Let $(x(\cdot,\cdot),\zeta(\cdot,\cdot),\rho(\cdot,\cdot))$  denote a  solution corresponding to an arbitrary initial condition $(x_0,\zeta_0,\rho_0) \in  \R^n \times  \mathbb{X}$ and assume that $x(\cdot,\cdot)$ is bounded. Then it holds that $(h(x(t,j),\zeta(t,j),\rho(t,j))) \to 0$
    for $(t,j) \to \infty$. 
    \hfill $\lrcorner$
\end{theorem}
The proof follows the ideas in \cite[Lem. 2, Thm. 1]{braun_2024}.

\begin{pf}
 \underline{(i)} The first item follows from \Cref{lem:ExistenceSolution} in combination with the properties of the feedback law $u_{\cdot}(\cdot)$ and the Lyapunov function $V_\cdot(\cdot)$ in \Cref{as:lyapunov_function}.

    \noindent \underline{(ii)} Note that $d(\zeta(t,j)) -V_{\zeta(t,j)}(x(t,j))\geq 0$ for all $(t,j) \in \operatorname{dom}(\zeta)$ implies~\eqref{eq:avoidance_property} according to the definition of $d(\cdot)$ in~\eqref{eq:def_d}. 
    Additionally, note that $d(\zeta) -V_{\zeta}(x)$ remains constant during jumps according to the discrete-time updates of $x$ and $\zeta$ in \eqref{eq:overall_hybrid_dynamics}.
    For the sake of a contradiction, assume that there exists a time $(T,J)\in \operatorname{dom}(x)$ such that 
    ${d(\zeta(T,J)) -V_{\zeta(T,J)}(x(T,J))< 0}$. This implies the existence of $(t_1,j_1)\in \operatorname{dom}(\zeta)$ such that $d(\zeta(t_1,j_1)) -V_{\zeta(t_1,j_1)}(x(t_1,j_1)) = 0$ and $(x(t_1,j_1),\zeta(t_1,j_1),\rho(t_1,j_1))\in \R^p \times \mathcal{C}$. 
    For any $(t_1,j_1) \in \dom(\zeta)$ with this property it holds that    
    $\dot{\zeta}=0$ and
    $$\langle \nabla V_{\zeta(t_1,j_1)} (x(t_1,j_1)),f(x(t_1,j_1),u_{\zeta(t_1,j_1)}(x(t_1,j_1)) \rangle < 0$$ according to \eqref{eq:lyapunov_condition}, i.e., $d(\zeta(t_1,j_1)) -V_{\zeta(t_1,j_1)}(x(t_1,j_1))$ is increasing leading to a contradiction.   
    
    \noindent \underline{(iii)} 
    Note that $d(\zeta)>0$ for all $(\zeta,\rho)\in \mathbb{X}$ according to the assumptions in the theorem and the definition of 
    $\mathbb{X}$ in \eqref{eq:def_X}.
    First, let $(x_0,\zeta_0,\rho_0)\in \R^n \times \mathbb{X}$ such that $d(\zeta_0) -V_{\zeta_0}(x_0) < 0$.  
    Thus, $\zeta(\cdot,\cdot)$ remains constant until the condition 
    \begin{align}
    d(\zeta(t,0)) -V_{\zeta(t,0)}(x(t,0)) >0        \label{eq:iii_proof_conditions}
    \end{align}
   is satisfied. 
According to \eqref{eq:convergence_z}, $\|x(t,0)\|_{G_x(\zeta_0)}$ becomes arbitrarily small for $t\rightarrow \infty$. Thus, there exists $(t,j) \in \dom(x)$ such that \eqref{eq:iii_proof_conditions} holds. Hence, it follows from item~(ii) that there exists $(T,J)\in \operatorname{dom}(x)$ such that
    $d(\zeta(t,j)) -V_{\zeta(t,j)}(x(t,j)) \geq 0$ 
    for all $(t,j)\geq (T,J)$, $(t,j)\in \operatorname{dom}(x)$.
Without the scaling $\max\{0, \ell (d(\zeta)-V_{\zeta}(x))\}\geq 0$ in \eqref{eq:overall_hybrid_dynamics} we know that the $\zeta$-dynamics converge to $0$ in finite-time (see \Cref{Thm:PathGlobal}). 
For the sake of a contradiction, assume that  $\zeta(\cdot,\cdot)$ as a solution of \eqref{eq:overall_hybrid_dynamics} does not converge to zero.
Hence, recalling Remark \ref{rem:scaling}, which states that $\zeta(\cdot,\cdot)$ as solutions of \eqref{eq:overall_hybrid_dynamics} corresponds to a time-scaling of  $\xi(\cdot,\cdot)$ as solutions of \eqref{eq:HybridPath}, there needs to exist $\zeta^\sharp \in \R^p$ such that
$\zeta(t,j) \rightarrow \zeta^\sharp$ for $(t,j) \rightarrow \infty$, $(t,j)\in \operatorname{dom}(\zeta)$.
Using the same steps an in the proof of \cite[Thm.~1]{braun_2024a}, for almost all $t\in \R_{\geq 0}$, $(t,j) \in \dom(x)$, it holds that\footnote{The time argument has been removed to shorten expressions.}
\begin{align}
\tfrac{d}{dt} V_{\zeta^\sharp}(x)  
&\leq \|\nabla V_{\zeta^\sharp}(x)\| \cdot \|f(x,u_{\zeta}(x))-f(x,u_{\zeta^\sharp}(x)) \| \nonumber \\
&\quad -\alpha_3(V_{\zeta^\sharp}(x)) \label{eq:ISS_ineq1}
\end{align}
where $\nabla V_\cdot(\cdot)$ denotes the gradient of $V_\cdot(\cdot)$ with respect to $x$.
Since $x(\cdot,\cdot)$ is bounded by assumption, there exists a compact set $\mathcal{C}_x\subset \R^n$ such that $x(t,j)\in \mathcal{C}_x$ for all $(t,j)\in \operatorname{dom}(x)$. Moreover, since $\zeta_{0}$ and $\zeta^\sharp$ are fixed, there exists a compact set $\mathcal{C}_{{\zeta}} \subset \R^p$ such that $\zeta(t,j) \in \mathcal{C}_\zeta$ for all $(t,j)\in \operatorname{dom}(\zeta)$.
Since $V_{\zeta^\sharp}(\cdot)$ is continuous and $f$ and $u_{\cdot}(\cdot)$ are locally Lipschitz continuous by assumption, there exist $M_V$ and $L_f$ such that 
$\|\nabla V_{\zeta^\sharp}(x)\| \leq M_V$ for all $x \in \mathcal{C}_x$ 
and
\begin{align*}
\|f(x,u_{\zeta_1}(x))-f(x,u_{\zeta_2}(x)) \| \leq L_f \|\zeta_1-\zeta_2\| 
\end{align*}
for all $\zeta_1,\zeta_2\in \mathcal{C}_\zeta$, for all $x\in \mathcal{C}_x$.
Using inequality
\eqref{eq:lyapunov_condition1} (i.e., $\alpha_1(\|x\|_{G_x(\zeta^\sharp)}) \leq V_{\zeta^\sharp}(x)$),
it holds that
\begin{align*}
\tfrac{d}{dt} V_{\zeta^\sharp}(x) &\leq -\alpha_3(V_{\zeta^\sharp}(x)) + M_V L_f \| \zeta - \zeta^\sharp \|   \\
&\leq -\tfrac{1}{2}\alpha_3(\alpha_1(\|x\|_{G_x(\zeta^\sharp)}) \\
& \qquad -\tfrac{1}{2}\alpha_3(\alpha_1(\|x\|_{G_x(\zeta^\sharp)}) + M_V L_f \| \zeta - \zeta^\sharp\|. 
\end{align*}
Next, we define $ \alpha_4(\cdot)=\alpha_1^{-1}(\alpha_3^{-1}(2M_V L_f(\cdot))) \in \mathcal{K}_{\infty}$.
Then,
    $\|x\|_{G_x(\zeta^\sharp)} \geq \alpha_4(\|\zeta - \zeta^\sharp\|)$
implies
\begin{align*}
\tfrac{d}{dt} V_{\zeta^\sharp}(x) 
\leq -\tfrac{1}{2}\alpha_3(\alpha_1(\|x\|_{G_x(\zeta^\sharp)}) 
\end{align*}
and thus $V_{\zeta^\sharp}(\cdot)$ is an ISS-Lyapunov function \cite[Sec. 2.2]{LIN19961954}.
There exist $\beta\in \mathcal{KL}$ and $\alpha_5\in \mathcal{K}_{\infty}$ so that solutions satisfy
\begin{align*}
    \|x(t,j)\|_{G_x(\zeta^\sharp)} \leq \beta(\|x(0,0)\|_{G_x(\zeta^\sharp)},t) + \alpha_5(\|\zeta - \zeta^\sharp\|_{L_{\infty}})
\end{align*}
for all $(t,j)\in \operatorname{dom}(x)$ and where $\|\cdot\|_{L_{\infty}}$ denotes the $L_\infty$-norm of a signal (see \cite[Prop. 3]{LIN19961954}).
Since $\zeta(t,j) \rightarrow \zeta^\sharp$ as $(t,j)\rightarrow \infty$, one concludes that
$\|x(t,j)\|_{G_x(\zeta^\sharp)}\rightarrow \zeta^\sharp$ for $(t,j)\rightarrow \infty$, i.e., $h(x(t,j))\rightarrow \zeta^\sharp$.
Since $d(\cdot)$ defined in \eqref{eq:def_d} is continuous  
and continuous functions attain their minimum on compact sets, we can define 
$\varepsilon_d \in ( 0, \min_{(t,j)\in\dom(\zeta)} d(\zeta(t,j)) )$.
Since $\zeta(t,j) \rightarrow \zeta^\sharp$ as $(t,j)\rightarrow\infty$, there exists $(t_1,j_1)\in \operatorname{dom}(\zeta)$ such that $V_{\zeta^\sharp}(x(t,j))\leq \frac{1}{2} \varepsilon_d$ for all $(t,j)\geq (t_1,j_1)$, $(t,j) \in \operatorname{dom}(\zeta)$.
Thus,  $\max\{0, \ell (d(\zeta(t,j))-V_{\zeta_s(t,j)}(x(t,j)))\} \geq \frac{\ell}{2} \varepsilon_d$ for all $(t,j)\geq (t_1,j_1)$, $(t,j) \in \operatorname{dom}(\zeta)$ in \eqref{eq:overall_hybrid_dynamics}, which contradicts the assumption
$\zeta^\sharp\neq 0$ and completes the proof.
\hfill $\qed$
\end{pf}

\begin{remark}
To incorporate input constraints in the controller design, one can introduce an upper bound on $d(\cdot)$, ensuring that the size of the sublevel sets of the Lyapunov functions are sufficiently small. 
\hfill $\diamond$
\end{remark}

\section{Numerical illustrations} \label{Sec:Numerics}
We illustrate the results via a numerical example.
Consider a unicycle mobile robot governed by the nonlinear dynamics
\begin{equation} \label{eq:UniCycleDynamics}
    \begin{aligned}
        \ddt \begin{smallbmatrix}
            p_1 \\ p_2 \\ \theta \\ w_1 \\ w_2
        \end{smallbmatrix} = 
        \begin{smallbmatrix}
            w_1 \cos(\theta) \\ w_1 \sin(\theta) \\ w_2 \\
            u_1 \\ u_2
        \end{smallbmatrix}, \quad z 
        = \begin{smallbmatrix}
            p_1 \\ p_2
        \end{smallbmatrix},
    \end{aligned}
\end{equation}
where
$x=(p_1,p_2,\theta,w_1,w_2) \in \R^{5}$ is the state,
$z=(x_1, x_2)$ is the output of interest, and
$u=(u_1,u_2) \in \R^2$ is the control input.
For $P := \begin{smallbmatrix} 1&1\\1&2 \end{smallbmatrix}$,
a Lyapunov function according and a corresponding feedback law satisfying \Cref{as:lyapunov_function} are given by
\begin{align} \label{eq:ExampleLyapunovFunction}
V_{z_e}(x) &:= V_{z_e}^1(z,\theta) +V_{z_e}^2(x) \\
V_{z_e}^1(z,\theta) &:= 
    \tfrac{1}{2} \langle \bar  p, P \bar p \rangle + \tfrac{1}{4} \left( \bar p_1^4 + \bar  p_2^4 \right), \nonumber \\
    V_{z_e}^2(x)&:=  \tfrac{1}{2} (w_1 - v_1(\bar{p}))^2 + \tfrac{1}{2} (w_2-v_2(\bar{p}))^2,  \nonumber
\end{align}
and
\begin{align}
    &\bar p \!:=\! \begin{smallbmatrix}
    \cos(\theta) & -\sin(\theta) \\ \sin(\theta) & \cos(\theta)
\end{smallbmatrix}^\top \!\! (z_e\! -\! p),  \ 
    v (\bar p) \!:=\! \begin{smallpmatrix} \!
        20 \bar p_1 \bar p_2^2 + 25 \bar p_1^3 + 20 \bar p_2^3 \\ 20 \bar p_1 \bar p_2 \!
    \end{smallpmatrix}, \nonumber \\
&u_{z_e,1}(x) = -w_1+v_{1}(\bar{p}) + \left( 75 \bar{p}_1^2  +20 \bar{p}_2^2\right) (-w_1+ w_2 \bar{p}_2) \nonumber \\
&\qquad \qquad  -\left( 60 \bar{p}_2^2 +40 \bar{p}_1 \bar{p}_2  \right) w_2 \bar{p}_1, \label{eq:ExampleInput} \\
&u_{z_e,2}(x)=-w_2+v_{2}(\bar{p})+20 \bar{p}_2 (- w_1+ w_2 \bar{p}_2) -20 \bar{p}_1^2 w_2. \nonumber
\end{align}
The underlying derivations can be found in \cite{braun_2024,braun_2024a}.
For the calculation of $d$ in \eqref{eq:def_d} we 
note that
\begin{align*}
    \inf_{\substack{z\in \cup_{i=1}^N \overline{B}_{r_i}(q_i)\\ z=h(x), \ x\in \R^n}} V_{z_e}(x) =     \inf_{\substack{z\in \cup_{i=1}^N \overline{B}_{r_i}(q_i)\\  \theta\in \R}} V_{z_e}^1(z,\theta) 
\end{align*}
which follows from the fact that for all $z\in \cup_{i=1}^N \overline{B}_{r_i}(q_i)$, $z=h(x)$, there exists $w_1,w_2 \in \R$ such that $V_{z_e}^2(x)=0$. As a next step we use the estimates $\lambda_{\min} \|z_e-z\|^2 \le \langle \bar p, P \bar p \rangle $,
where $\lambda_{\min} > 0$ denotes the smallest eigenvalue of~$P$  
and $\| z_e - z\|^4 \le 2(\bar p_1^4 + \bar p_2^4)$ 
to define
\begin{align*}
d(z_e):=&\inf_{\substack{z\in \cup_{i=1}^N \overline{B}_{r_i}(q_i)}} \tfrac{\lambda_{\min}}{2} \|z_e-z\|^2 + \tfrac{1}{8}\|z_e-z\|^4  \\
\leq &   \inf_{\substack{z\in \cup_{i=1}^N \overline{B}_{r_i}(q_i) \\ \theta\in \R}} V_{z_e}^1(z,\theta) 
\end{align*}
as a selection of $d(\cdot)$ defined in \eqref{eq:def_d}.\footnote{For the simulation 
we use $\tilde \varepsilon=0$, which only guarantees that the interior of each obstacle is avoided. To additionally ensure avoidance of the boundary, an arbitrarily small parameter $\tilde \varepsilon>0$ can be used instead. As in \cite{braun_2024}, $d(z_e)$ is calculated by discretizing the boundary of the obstacles.}
With feedback controller~\eqref{eq:ExampleInput} it was shown in \cite[Sec.~IV-C]{braun_2024} that the unicycle~\eqref{eq:UniCycleDynamics} follows a pre-calculated path through a cluttered environment with various obstacles.
We use the same feedback while we generate a safe path online using~\eqref{eq:HybridFeedbackLaw},~\eqref{eq:HybridPath} and without knowing the obstacles' positions in advance.
\begin{figure}
    \centering
    \includegraphics[width=1\columnwidth]{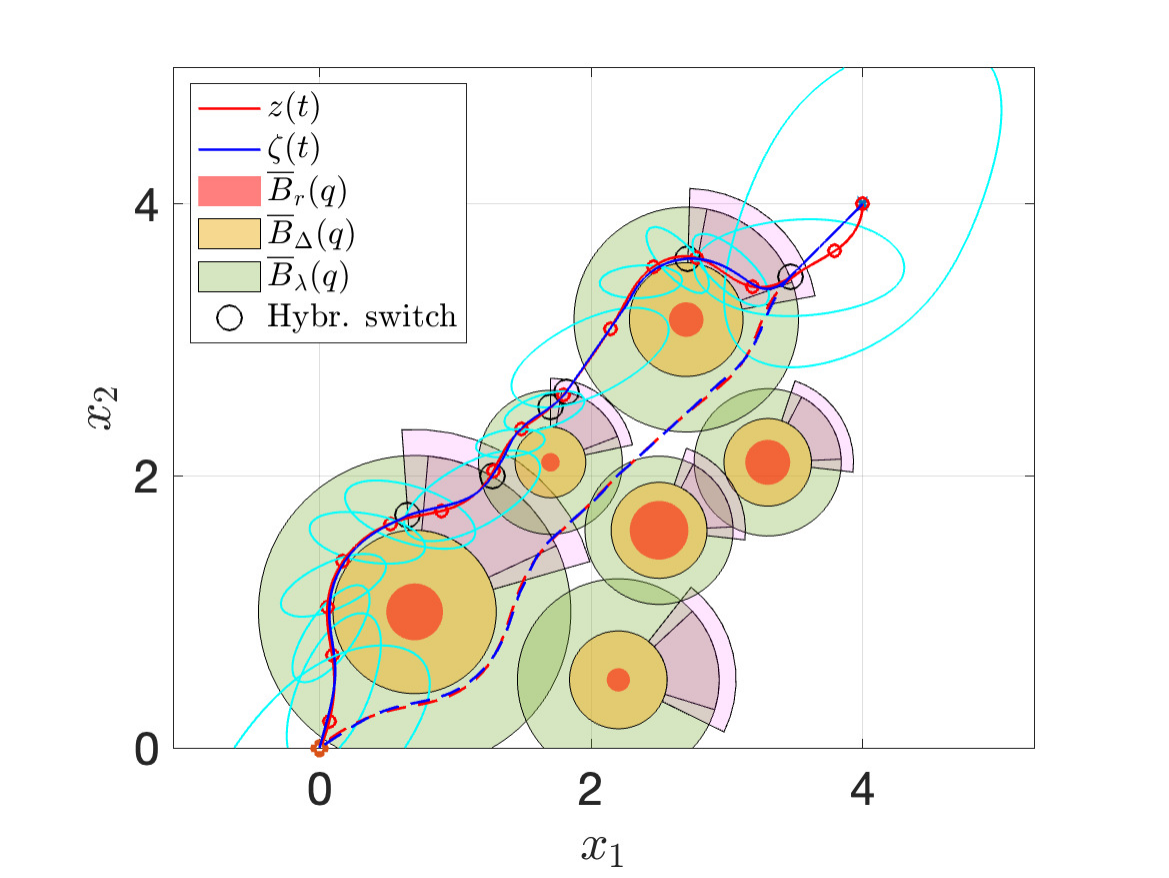}
    
    \vspace{-0.1cm}
    
    \includegraphics[width=0.9\columnwidth]{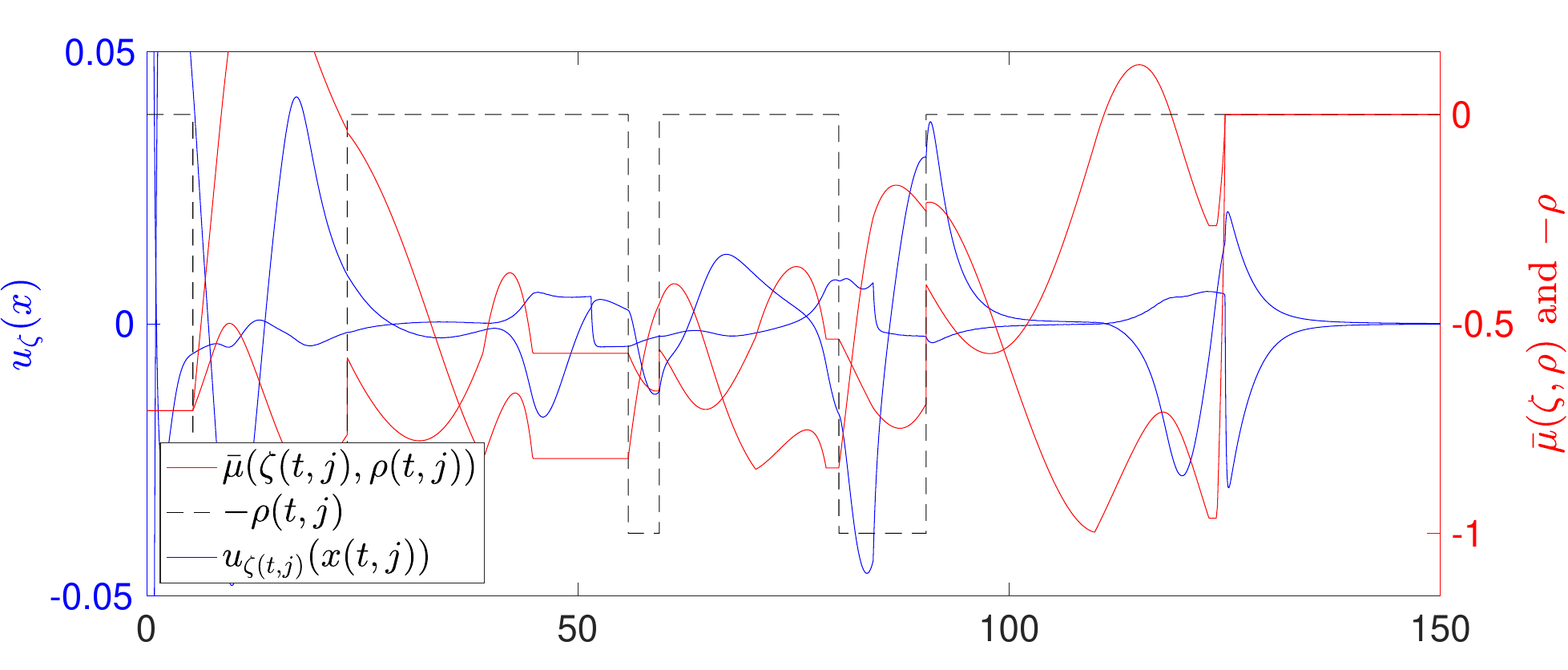}
    \caption{Top: Simulation of system~\eqref{eq:UniCycleDynamics}, feedback law~\eqref{eq:ExampleInput} and paths given by~\eqref{eq:HybridPath}.
    Magenta areas refer to the sets~$M_0^i, M_1^i$, cyan lines represent level sets of Lyapunov functions~\eqref{eq:ExampleLyapunovFunction} centered at the small red circles, and
    solid and dashed lines result from different choices of~$\bar q_i$ in~\eqref{eq:HybridFeedbackLaw}.
    Bottom: Time evolution of the input~$u_{\zeta(t,j)}(x(t,j))$, the virtual input~$\bar \mu(\zeta(t,j),\rho(t,j))$, and the hybrid state~$-\rho(t,j)$.
    }
    \label{fig:Simulation_UniCycle}
\end{figure}
\Cref{fig:Simulation_UniCycle} (top) shows the evolution in the~$(x_1,x_2)$-plane.
The blue line is the safe path~$\zeta(t,j)$ from~\eqref{eq:HybridFeedbackLaw},~\eqref{eq:HybridPath}. 
The red line is~$z(t)$ from~\eqref{eq:UniCycleDynamics} using~\eqref{eq:ExampleInput} to follow the path.
A red area represents obstacle~$i$ with radius~$r_i > 0$, the orange area is the safety zone (the path is not allowed to enter but the system is) with radius~$\Delta_i = r_i + \delta_i$, and the green area is the activation zone with radius~$\lambda_i > \Delta_i$, respectively.
Magenta areas represent the sets~$M_0^i, M_1^i$, and the small circles refer to the jumps of the state~$\rho$ determining if~$\eqref{eq:ControllerNomalMode}$ or~\eqref{eq:ControllerHybridMode} is active.
\Cref{fig:Simulation_UniCycle} (bottom) shows the evolution of the input~$u_{\zeta(t,j)}(x(t,j))$
    from~\eqref{eq:ExampleInput}, the virtual input~$\bar \mu(\zeta(t,j),\rho(t,j))$
   from~\eqref{eq:HybridFeedbackLaw}, and the hybrid state~$-\rho(t,j)$.

\section{Conclusions and outlook} \label{sec:conclusions}

We have proposed a solution to the reach-and-avoid Problem \ref{prob:problem_formulation} for 
dynamical systems achieving obstacle avoidance of circular obstacles and convergence to the origin of the output of interest. The overall 
design merges a controller for simple virtual dynamics generating a reference signal with a Lyapunov based controller for the original dynamics. The 
results rely on the simplicity of the virtual dynamics and on forward invariance of Lyapunov sublevel sets. While circular obstacles  
seem restrictive, future work will focus on the coordination of a fleet of mobile robots where each robot represents a spherical moving obstacle that needs to be guided to a target position.

\end{document}